\documentclass[twocolumn,prb,showpacs]{revtex4}
\usepackage{graphicx}
\usepackage{dcolumn}
\usepackage{bm}
\usepackage{amsmath}

\begin{document}

\preprint{}
\title{ Anomalous Josephson Hall effect in magnet/triplet superconductor junctions}
\author{Takehito Yokoyama}
\affiliation{Department of Physics, Tokyo Institute of Technology, Tokyo 152-8551,
Japan 
%\\ $^2$Department of Physics, Tokyo Metropolitan University, Hachioji, Tokyo 192-0397, Japan 
}
\date{\today}

\begin{abstract}
We investigate anomalous Hall effect in a magnet coupled to a triplet superconductor under phase gradient. It is found that the anomalous Hall supercurrent arises from non-trivial structure of the magnetization. The magnetic structure manifested in the Hall supercurrent is characterized by even order terms of the exchange coupling, essentially different from that discussed in the context of anomalous Hall effect, reflecting the disspationless nature of supercurrent. We also discuss a possible candidate for magnetic structure to verify our prediction.

\end{abstract}

\pacs{73.43.Nq, 72.25.Dc, 85.75.-d}
\maketitle

%\affiliation{$^1$Department of Physics, Tokyo Institute of Technology, 2-12-1 Ookayama, Meguro-ku, Tokyo 152-8551, Japan \\

Recently, the coexistence of superconductivity and ferromagnetism has received much attention.\cite{Buzdin,Bergeret,Eschrig,Linder} In ferromagnet/superconductor junctions, equal-spin triplet pairings can be generated due to spin flip scattering in ferromagnetic multilayer or inhomogeneous ferromagnet\cite{Bergeret2}, or in uniform ferromagnet with spin-orbit coupling\cite{Bergeret3}. Spin-polarized supercurrent carried by equal-spin triplet pairings is a novel ingredient for spintronics applications. Recent experiments have successfully demonstrated the generation of spin-triplet pairing by observing Josephson current through strong ferromagnets.\cite{Keizer,Khaire,Robinson}  
Up to now, in ferromagnetic Josephson junctions, longitudinal Josephson current has been investigated.\cite{Bergeret4,Golubov}

Ferromagnet/triplet superconductor junctions offer richer physics than singlet ones since triplet pairings have spin degree of freedom. The interaction between magnetism and triplet superconductivity leads to interesting phenomena such as 0-$\pi$ transitions.\cite{Kastening,Brydon,Brydon2,Brydon3,Yokoyama,Gentile} 
%By rotating the spin of the triplet pairings, 0-$\pi$ transition can be induced. In fact, it has been shown that, in ferromagnetic Josephson junctions with triplet superconductors, the interplay between triplet superconductivity and magnetism is manifested in the Josephson current. 
Josephson junctions composed of triplet superconductors have been fabricated to identify the pairing symmetry of triplet superconductors.\cite{Nelson,Strand,Strand2} Also, a highly conducting interface between a ferromagnet SrRuO$_3$ and a triplet superconductor Sr$_2$RuO$_4$ has been realized recently.\cite{Anwar}

The Hall effect in ferromagnets has been discussed intensively in the context of anomalous Hall effect. \cite{Nagaosa}
The anomalous Hall effect arises from non-trivial spin texture, which is associated with the spin Berry phase effect. \cite{Ye,Taguchi,Ohgushi,Tatara,Onoda}
It has been shown that the Hall conductivity stems from non-trivial spin configurations such as vector spin chirality ${\bf{S}}_i  \times {\bf{S}}_j$ \cite{Taguchi2} and scalar spin chirality  ${\bf{S}}_i  \cdot ({\bf{S}}_j  \times {\bf{S}}_k )$ \cite{Tatara}, where ${\bf{S}}_i$ is a localized spin with position $i$. Motivated by these studies, in this paper, we investigate Hall supercurrent driven by non-trivial magnetic structure coupled to triplet superconductivity under phase gradient. Since phase is odd in time-reversal, the magnetic structure manifested in Hall supercurrent becomes essentially different from that in the anomalous Hall effect.

In this paper, we study the anomalous Hall effect in a magnet coupled to a triplet superconductor under phase gradient. It is shown that the anomalous Hall supercurrent arises from non-trivial structure of the magnetization. The magnetic structure manifested in the Hall supercurrent is essentially different from that discussed in the context of anomalous Hall effect, reflecting the disspationless nature of supercurrent. 
We also discuss the condition of the generation of the anomalous Josephson Hall current based on symmetry and a possible candidate for magnetic structure to verify our prediction.

\begin{figure}[tbp]
\begin{center}
\scalebox{0.8}{
\includegraphics[width=8.50cm,clip]{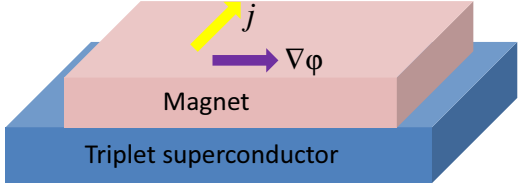}
}
\end{center}
\caption{(Color online) Schematic illustration of magnet/triplet superconductor junction.}
\label{fig1}
\end{figure}

%%%%%%%%%%%%%%%%%%%%% Formulation
We consider a magnet/triplet superconductor junction (See  Fig. \ref{fig1}).
The Hamiltonian of the superconductor and the magnet are given by  $H_S  = H_0  + H_\Delta $ and $H_M  = H_0  + H_{ex} + H_\varphi$, respectively. 
The $H_0$, $H_\Delta$ and  $H_{ex}$ represent the kinetic energy, the superconducting order parameter, and the exchange interaction between the conducting electrons and the local spins, respectively:
\begin{eqnarray}
H_0  = \sum\limits_{\bf{k}} {\phi _{\bf{k}}^\dag  \xi {\sigma _0  \otimes } \tau _3 \phi _{\bf{k}}^{} }  ,\\ H_\Delta   = \sum\limits_{\bf{k}} {\phi _{\bf{k}}^\dag  {({\bf{d}}  \cdot {\bm{\sigma }}) \otimes \tau_1 }}, \\ 
H_{ex}  =  - J\sum\limits_{{\bf{k}},{\bf{q}}} {(\phi _{{\bf{k}} - {\bf{q}}}^\dag  {\bm{\sigma }} \otimes \tau _0 \phi _{\bf{k}}^{} ) \cdot {\bf{S}}_{\bf{q}}^{} } \label{hex}
\end{eqnarray}
with $\phi _{\bf{k}}^\dag   = (c_{{\bf{k}} \uparrow }^\dag  ,c_{{\bf{k}} \downarrow }^\dag  ,ic_{ - {\bf{k}} \downarrow }^{} , - ic_{ - {\bf{k}} \uparrow }^{} )$ and $\xi  = \frac{{\hbar ^2 k^2 }}{{2m}} - \varepsilon _F$. Here, $\sigma$ and $\tau$ denote the Pauli matrices in spin and Nambu spaces, respectively. Also, $\varepsilon _F$, ${\bf{d}}$, $J$, and $\bf{S}$ are the Fermi energy, $\bf{d}$-vector of the triplet pairing, the exchange coupling, and the localized spins, respectively. 
The localized spins can have spatial dependence, but we consider only slowly varying case. 
Note that we adopt the basis in Ref.\cite{Ivanov} such that singlet pairing is proportional to the unit matrix in spin space. 
We consider supercurrent induced by phase gradient.
The phase gradient along $j$ direction, $\nabla _j \varphi$, enters the Hamiltonian as 
\begin{eqnarray}
H_\varphi   = \sum\limits_{\bf{k}} {\phi _{\bf{k}}^\dag  \frac{{\hbar^2}}{m}k_j \nabla _j \varphi \phi _{\bf{k}}^{} } 
\end{eqnarray}
where
% $\varphi $ is half the phase of superconducting correlation and 
$\nabla _j \varphi$ is assumed to be spatially constant. 
We will treat $H_{ex}$ and $H_\varphi$ perturbatively. 

With the above Hamiltonians,
% the velocity operator reads 
%\begin{eqnarray}
%v_i  = \frac{\hbar }{m}k_i \tau _3  + \delta _{ij} \frac{e}{m}\nabla _j \varph\end{eqnarray}
the charge current operator ($j_{c}$) in $i$-direction is given by 
\begin{eqnarray}
j_{c,i}^{}  =  - \frac{{e\hbar }}{m}k_i \sigma _0  \otimes \tau _0  - \delta _{ij} \frac{{e\hbar  }}{m}\nabla _j \varphi \sigma _0  \otimes \tau _3
\end{eqnarray}
where $-e$ is the electron charge.

Before proceeding to the explicit calculation, let us discuss Josephson Hall current qualitatively based on symmetry argument. \cite{Murakami}
Consider the London equation of the form:
\begin{equation}
{\bf{j}}_c  =  - \frac{{e^2 }}{m}\rho  \cdot {\bf{A}}.
\end{equation}
where ${\bf{j}}_c$, $\rho$, and ${\bf{A}}$ are, respectively,  the charge current, the superfluid density tensor, and  the vector potential. 
Since charge current and vector potential are time-reversal odd, $\rho$ describes the reversible and dissipationless flow of the supercurrent. Thus, the Hall current can flow without breaking time-reversal symmetry. Namely, the Hall current is allowed in even-order perturbation with respect to time-reversal breaking term $H_{ex}$. 
This contrasts with the anomalous Hall effect\cite{Tatara} where Hall current is driven by an electric field which is even under time-reversal. Thus, one can expect an essentially different magnetic structure manifested in the Hall supercurrent. 
Also, since charge current and vector potential are odd under spatial inversion, the Hall coefficient should be even with respect to spatial gradient. 
Therefore, in the lowest order, the possible form of the Hall coefficient should be ${\nabla _i}{\bf{S}} \times {\nabla _j}{\bf{S}}$ with $i \ne j$.
However, this is a vector quantity while the Hall coefficient is a scalar.
Here, triplet superconductivity plays an essential role: We have a vector formed by triplet pairing (such as $\bf{d}$-vector). This way, we expect that the Hall coefficient is proportional to ${\nabla _i}{\bf{S}} \times {\nabla _j}{\bf{S}}$ projected on some vector due to triplet superconductivity. 
Below, we will show that this consideration is in fact correct by the explicit calculation of the Hall supercurrent.

%%%%%%%%%%%%%%%%%%%%%%%%%%%%%%%%%%%%%%%%%%%%%%%%%%%%%%%%%%%%%%%%%%%%%%%%%%%%%%%%%%%%%

\begin{figure}[tbp]
\begin{center}
\scalebox{0.8}{
\includegraphics[width=7.50cm,clip]{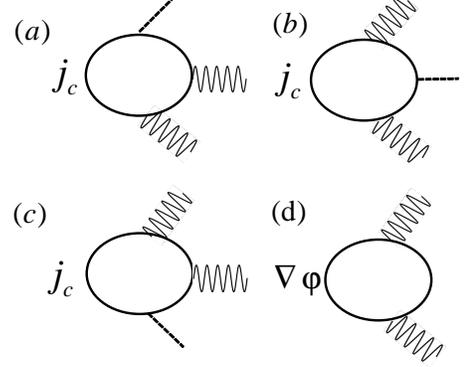}
}
\end{center}
\caption{ Diagrammatic representations of the current density. The wavy lines denote the interaction with the local spin $\bf{S}$ and dotted lines represent the phase gradient $\nabla \varphi$. (a-c) The contributions from the first term in Eq.(5). (d) The contribution from the second term in Eq.(5).}
\label{fig2}
\end{figure}

Now, we calculate Hall supercurrent and give their analytical expressions. We consider the unperturbed advanced Green's functions in the magnet of the form 
\begin{eqnarray}
g_{{\bf{k}},\omega }^a  = g_{0,{\bf{k}},\omega }^a \sigma _0  \otimes \tau _0  + g_{3,{\bf{k}},\omega }^a \sigma _0  \otimes \tau _3 + ({\bf{f}}_{{\bf{k}},\omega }^a  \cdot {\bm{\sigma }}) \otimes \tau _1 
\end{eqnarray}
where $g_{0,{\bf{k}},\omega }^a$ and $g_{3,{\bf{k}},\omega }^a$ are normal Green's functions while ${\bf{f}}_{{\bf{k}},\omega }^a$ is a 3D vector characterizing anomalous Green's function. The anomalous Green's function in the magnet arises due to the proximity effect.
We take into account $H_{ex}$ up to second order and $H_\varphi$ as a first order perturbation. 
Diagrammatic representations of the Hall supercurrent are shown in Fig. \ref{fig2}.

The Hall supercurrent can be represented as \cite{Haug}
\begin{eqnarray}
 j_{c,i}  =  \frac{{i\hbar ^2 e}}{{mV}}\sum\limits_{{\bf{k}},{\bf{q}}} {e^{ - i{\bf{q}} \cdot {\bf{x}}} {\rm{Tr}}k_i G_{{\bf{k}} - {\bf{q}}/2,{\bf{k}} + {\bf{q}}/2}^ <  (t,t)} \nonumber \\ 
  + \delta _{ij} \frac{{i\hbar^2  e}}{{mV}}\nabla _j \varphi \sum\limits_{{\bf{k}},{\bf{q}}} {e^{ - i{\bf{q}} \cdot {\bf{x}}} {\rm{Tr}}\tau _3 G_{{\bf{k}} - {\bf{q}}/2,{\bf{k}} + {\bf{q}}/2}^ <  (t,t)} 
\end{eqnarray}
where $V$ is the total volume and ${\rm{Tr}}$ is taken over spin and Nambu spaces. $G_{{\bf{k}} - {\bf{q}}/2,{\bf{k}} + {\bf{q}}/2}^ <  (t,t)$ is the lesser Green's function of the total Hamiltonian. 
Performing perturbations with respect to $H_{ex}$ and $H_\varphi$, we expand the lesser component using the advanced Green's functions by the Langreth theorem.\cite{Haug}
Noting that $g_{{\bf{k}},\omega }^ <   = f_\omega  \left[ {g_{{\bf{k}},\omega }^a  - (g_{{\bf{k}},\omega }^a )^\dag  } \right]$ with the lesser Green's function $g_{{\bf{k}},\omega }^ <$ and the Fermi distribution function $f_\omega$, and $\delta _{ij}  = \frac{{\partial k_i }}{{\partial k_j }}$, we can compute the Hall supercurrent.
Then, the Hall supercurrent ($i \ne j$) in the second orders in $J$ and spatial derivative is given by
\begin{widetext}
\begin{eqnarray}
j_{c,i}^{} = \frac{{16{\hbar ^4}e{J^2}}}{{{m^2}V}}{\nabla _j}\varphi \sum\limits_{{\bf{k}},\omega ,\alpha } {{{\left( {{\nabla _j}{{\bf{S}}^{}}({\bf{x}}) \times {\nabla _i}{{\bf{S}}^{}}({\bf{x}})} \right)}^\alpha }{f_\omega } } 
\nonumber \\
\times {\mathop{\rm Im}\nolimits} \left[ \begin{array}{l}
2{k_i}{k_j}\left( {\frac{\partial }{{\partial {k_i}}}\left\{ {{{(g_{0,{\bf{k}},\omega }^a)}^2} + {{(g_{3,{\bf{k}},\omega }^a)}^2}} \right\}{{\left( {{\bf{f}}_{{\bf{k}},\omega }^a \times \frac{\partial }{{\partial {k_j}}}{\bf{f}}_{{\bf{k}},\omega }^a} \right)}^\alpha } - \frac{\partial }{{\partial {k_j}}}\left\{ {{{(g_{0,{\bf{k}},\omega }^a)}^2} + {{(g_{3,{\bf{k}},\omega }^a)}^2}} \right\}{{\left( {{\bf{f}}_{{\bf{k}},\omega }^a \times \frac{\partial }{{\partial {k_i}}}{\bf{f}}_{{\bf{k}},\omega }^a} \right)}^\alpha }} \right)\\
 + {k_i}\left\{ {{{(g_{0,{\bf{k}},\omega }^a)}^2} + {{(g_{3,{\bf{k}},\omega }^a)}^2}} \right\}{\left( {{\bf{f}}_{{\bf{k}},\omega }^a \times \frac{\partial }{{\partial {k_i}}}{\bf{f}}_{{\bf{k}},\omega }^a} \right)^\alpha }
\end{array} \right].
\end{eqnarray}
Here, we have assumed that the proximity effect is weak: $\left| {{\bf{f}}_{{\bf{k}},\omega }^a} \right| \ll \left| {g_{0,{\bf{k}},\omega }^a} \right|,\left| {g_{3,{\bf{k}},\omega }^a} \right|$, and dropped the forth order terms in the anomalous Green's functions. We have also dropped the symmetric term ${\nabla _j}{{\bf{S}}^{}}({\bf{x}}) \cdot {\nabla _i}{{\bf{S}}^{}}({\bf{x}})$ since we are interested in the Hall effect.
%
%We see that $ \rho _c^{}$ depends only on junction parameters (namely, the unperturbed advanced Green's functions), independent of the details of the magnet. 
If the anomalous Green's function $f_{{\bf{k}},\omega }^a$ becomes zero, then the supercurrent becomes zero as expected. 
Assuming that the proximity effect is weak, we now consider the Green's functions of the form: 
$g_{0,{\bf{k}},\omega }^a + g_{3,{\bf{k}},\omega }^a = \frac{1}{{\omega  - i\gamma  - \xi }} \equiv g_{{\bf{k}},\omega }^a,\;g_{0,{\bf{k}},\omega }^a - g_{3,{\bf{k}},\omega }^a = \frac{1}{{\omega  - i\gamma  + \xi }} \equiv \bar g_{{\bf{k}},\omega }^a$, and $\;{\bf{f}}_{{\bf{k}},\omega }^a =g_{{\bf{k}},\omega }^a \bar g_{{\bf{k}},\omega }^a {\bf{d}}_{\bf{k}}$. Here, $\gamma$ is the scattering rate by impurities. 

Then, we have 
\begin{eqnarray}
j_{c,i}^{} = \frac{{8{\hbar ^4}e{J^2}}}{{{m^2}V}}{\nabla _j}\varphi \sum\limits_{{\bf{k}},\omega ,\alpha } {{{\left( {{\nabla _j}{{\bf{S}}^{}}({\bf{x}}) \times {\nabla _i}{{\bf{S}}^{}}({\bf{x}})} \right)}^\alpha }} {f_\omega }\nonumber \\ 
 \times {\mathop{\rm Im}\nolimits} \left[ \begin{array}{l}
\frac{{4{\hbar ^2}}}{m}k_i^2{k_j}\left\{ {{{(g_{{\bf{k}},\omega }^a)}^5}{{(\bar g_{{\bf{k}},\omega }^a)}^2} - {{(g_{{\bf{k}},\omega }^a)}^2}{{(\bar g_{{\bf{k}},\omega }^a)}^5}} \right\}{\left( {{\bf{d}}_{\bf{k}}^{} \times \frac{\partial }{{\partial {k_j}}}{\bf{d}}_{\bf{k}}^{}} \right)^\alpha } \\ - \frac{{4{\hbar ^2}}}{m}{k_i}k_j^2\left\{ {{{(g_{{\bf{k}},\omega }^a)}^5}{{(\bar g_{{\bf{k}},\omega }^a)}^2} - {{(g_{{\bf{k}},\omega }^a)}^2}{{(\bar g_{{\bf{k}},\omega }^a)}^5}} \right\}{\left( {{\bf{d}}_{\bf{k}}^{} \times \frac{\partial }{{\partial {k_i}}}{\bf{d}}_{\bf{k}}^{}} \right)^\alpha }\\ 
 + {k_i}\left\{ {{{(g_{{\bf{k}},\omega }^a)}^4}{{(\bar g_{{\bf{k}},\omega }^a)}^2} + {{(g_{{\bf{k}},\omega }^a)}^2}{{(\bar g_{{\bf{k}},\omega }^a)}^4}} \right\}{\left( {{\bf{d}}_{\bf{k}}^{} \times \frac{\partial }{{\partial {k_i}}}{\bf{d}}_{\bf{k}}^{}} \right)^\alpha }
\end{array} \right].
\end{eqnarray}
\end{widetext}

As an example of the ${\bf{d}}$-vector, consider the form of the Rashba type with the hexagonal warping:\cite{Fu}
\begin{eqnarray}
{\bf{d}}_{\bf{k}}^{} = \Delta _0^{}\left( {\begin{array}{*{20}{c}}
{{k_y}/{k_F}},&{ - {k_x}/{k_F}},&{\lambda (k_x^3 - 3k_x^{}k_y^2)/k_F^3}
\end{array}} \right)
\end{eqnarray}
where $k_F$ is the Fermi wavelength.
Then, under phase gradient in $x$-direction, we obtain the Hall supercurrent in $y$-direction driven by magnetic structure:
\begin{eqnarray}
j_{c,y}^{} =  - \frac{{38e{J^2}\Delta _0^2\lambda \nu }}{{105{\hbar ^4}k_F^4\varepsilon _F^2}}{\left( {{\nabla _x}{{\bf{S}}^{}}({\bf{x}}) \times {\nabla _y}{{\bf{S}}^{}}({\bf{x}})} \right)^x}{\nabla _x}\varphi \label{jy}
\end{eqnarray}
in the limit of $\gamma \to 0$. Here, $\nu$ is the density of states at the Fermi level.

This is in contrast to the normal Hall current in ferromagnet:
In the normal state, the Hall current is driven by scalar spin chirality under electric field \cite{Tatara}
\begin{eqnarray}
j_{c,y}  \propto \left( {{\nabla _x}{{\bf{S}}^{}}({\bf{x}}) \times {\nabla _y}{{\bf{S}}^{}}({\bf{x}})} \right) \cdot {\bf{S}}({\bf{x}}). \label{jcn}
\end{eqnarray}
By comparing Eq.(\ref{jy}) and Eq.(\ref{jcn}), we find an essentially different magnetic structure of Hall supercurrent, which reflects that supercurrent flows in response to phase gradient, i.e. reflects the disspationless nature of supercurrent.
Note that the normal Hall current is driven in an electric field which is even under time reversal. Hence, the Hall coefficient in the normal state is composed of odd order terms with respect to ${\bf{S}}$.

Now, we discuss a possible candidate of the magnetic structure that may be used to verify our prediction. First, the magnetization vector ${\bf{S}}(\bf{x})$ should have both $x$ and $y$ dependences. To observe finite Hall supercurrent, $\nabla_x {\bf{S}}(\bf{x})$ and $\nabla_y {\bf{S}}(\bf{x})$ should not be parallel to each other. 
One possible candidate is a magnetic skyrmion in chiral magnets\cite{Rossler,Muhlbauer,Jonietz,Nagaosa2,Fert}, such as MnSi, characterized by
\begin{eqnarray}
{{\bf{S}}^{}}({\bf{x}}) = \frac{S}{{{x^2} + {y^2} + {\Lambda ^2}}}\left( {\begin{array}{*{20}{c}}
{ - 2\Lambda x,}&{2\Lambda y,}&{{x^2} + {y^2} - {\Lambda ^2}}
\end{array}} \right)
\end{eqnarray}
where $\Lambda$ determines the size of the skyrmion.
For this spin texture, we have 
\begin{eqnarray}
{\left( {{\nabla _x}{{\bf{S}}^{}}({\bf{x}}) \times {\nabla _y}{{\bf{S}}^{}}({\bf{x}})} \right)^x} = \frac{{8{S^2}{\Lambda ^3}y}}{{{{({x^2} + {y^2} + {\Lambda ^2})}^3}}}.
\end{eqnarray}
Since this is odd in $y$, the total Hall supercurrent vanishes for infinite systems or when the skyrmion is located in symmetric points with respect to $y$-coordinate in finite systems. Note, however, that the skyrmion can be moved by injecting ultralow current.\cite{Jonietz,Fert}
For example, consider the skyrmion located in the region $- \infty  < x < \infty , - a \le y \le b$. Then, the total Hall supercurrent along the edges is proportional to
\begin{widetext}
\begin{eqnarray}
\int_{ - \infty }^\infty  {dx{{\left. {{{\left( {{\nabla _x}{{\bf{S}}^{}}({\bf{x}}) \times {\nabla _y}{{\bf{S}}^{}}({\bf{x}})} \right)}^x}} \right|}_{y = b}} + {{\left. {{{\left( {{\nabla _x}{{\bf{S}}^{}}({\bf{x}}) \times {\nabla _y}{{\bf{S}}^{}}({\bf{x}})} \right)}^x}} \right|}_{y =  - a}}}  = \frac{{32{S^2}{\Lambda ^3}}}{3}\left[ {\frac{b}{{{{({b^2} + {\Lambda ^2})}^{5/2}}}} - \frac{a}{{{{({a^2} + {\Lambda ^2})}^{5/2}}}}} \right]
\end{eqnarray}
\end{widetext}
which becomes $\frac{{32{S^2}{\Lambda ^3}}}{3}\left( {\frac{1}{{{b^4}}} - \frac{1}{{{a^4}}}} \right)$ for $\Lambda  \ll a,b$.

We also discuss possible forms of the ${\bf{d}}$-vector and magnetic structures from the symmetry of the Hamiltonian. Consider quasi two dimensional systems in $xy$-plane, focusing on the magnetic region with proximity induced triplet superconductivity, and the current along the $y$-direction under the phase gradient along the $x$-axis. Since the current density is transformed as $j_{c,y}(x,y) \to -j_{c,y}(-x,y)$ under the operation of $C_{2y} T$, where $T$ denotes time-reversal, if the Hamiltonian is invariant under this operation, $j_{c,y}$ is an odd function of $x$. Thus, if the Hamiltonian has this symmetry, the net Hall supercurrent becomes zero. This symmetry requires the condition ${d_x}({k_x},{k_y}) = d_x^*({k_x}, - {k_y}),  {d_y}({k_x},{k_y}) =  - d_y^*({k_x}, - {k_y})$, and ${d_z}({k_x},{k_y}) = d_z^*({k_x}, - {k_y})$ where ${\bf{d}}_{\bf{k}} = (d_x, d_y, d_z)$. 
Therefore, this symmetry is broken in the ${\bf{d}}$-vector of Eq.(11), which is consistent with the above results of the Hall supercurrent. The skyrmion magnetic structure in Eq.(14) respects this symmetry. 
Thus, if we consider an $s$-wave singlet pairing instead of triplet one, since the singlet pairing respects the symmetry, the Hall suppercurrent vanishes.

Let us estimate the Hall supercurrent  using Eq. (\ref{jy}).
For $\varepsilon _F \sim 10$ eV, $J \sim $ 100 meV, $\nu \sim 0.1$ /eV/unit cell, $\Delta_0 \sim$ 0.01 meV, $\lambda \sim 0.1$, $k_F \sim 10^{10}$m$^{-1}$, ${{\nabla _x}{{\bf{S}}^{}}({\bf{x}}) \times {\nabla _y}{{\bf{S}}^{}}({\bf{x}})} \sim 10^{16}$m$^{-2}$, $\nabla _x \varphi  \sim$ (100 nm)$^{ - 1}$, and the lattice constant $\sim 5$ \AA,  we estimate the magnitude of the current as $j_{c,y} \sim 5 \times 10^{5}$ A/cm$^2$ which is of measurable magnitude\cite{Oboznov}.

In this paper, we have considered magnet/triplet superconductor junctions. Since ferromagnets with intrinsic spin-orbit coupling can generate triplet pairings,\cite{Bergeret3} one can instead consider magnet/singlet superconductor junctions with intrinsic spin-orbit coupling in the magnet. Also, since inhomogeneous magnetism can convert singlet pairing into triplet one,\cite{Bergeret2} magnet/magnet/singlet superconductor junctions such as MnSi/Ho/Nb junctions (Ho is a conical magnet) can be used.
One may also consider MnSi/Bi$_2$Te$_3$ film/Nb junctions. Here, due to the proximity effect, triplet pairing like Eq.(11) can be induced in the Bi$_2$Te$_3$ film. This induced pairing can further penetrate into MnSi.

Spin Hall effect due to Rashba type spin-orbit coupling in superconductor~\cite{Kontani} or Josephson junctions~\cite{Mal'shukov} has been discussed.
In this paper, we have predicted Hall supercurrent driven by non-trivial magnetic texture, and hence our results do not rely on spin-orbit coupling. 
%Since spin current is even under time-reversal while electric field is odd,  time-reversal symmetry should be broken to obtain finite spin Hall current under electric field. 
In Ref.~\cite{Mal'shukov}, spin Hall effect is obtained by applying electric bias to the Josephson junction in order to make the Josephson current time dependent. 
In sharp contrast, we have considered stationary supercurrent under non-trivial magnetic structure when phase gradient is applied.
%, which is odd under time-reversal,

%In this paper, we have predicted the presence of off-diagonal component of the superfluid density tensor in ferromagnetic junctions. The anisotropy of superfluid density tensor also appears in anisotropic superlfuids in $^3$He A-phase.\cite{Volovik} However, the mechanisms are completely different: the Hall supercurrent arises from the inhomogeneous magnetization. On the other hand, in $^3$He A-phase, the anisotropic superfluid density stems from $l$-texture (inhomogeneous $l$-vector). 

In summary,
we have investigated anomalous Hall effect in a magnet coupled to a triplet superconductor under phase gradient. It is found that the anomalous Hall supercurrent arises from non-trivial structure of the magnetization. The magnetic structure manifested in the Hall supercurrent is essentially different from that discussed in the context of anomalous Hall effect, reflecting the disspationless nature of supercurrent. We have also discussed the condition of the generation of the anomalous Josephson Hall current based on symmetry and a possible candidate of magnetic structure to verify our prediction.

%This work was supported by Grant-in-Aid for Young Scientists (B) (No. 23740236) , 
%the MEXT KAKENHI (No. 23103505). 
%the "Topological Quantum Phenomena" (No. 25103709) Grant-in Aid for Scientific Research on Innovative Areas from the Ministry of Education, Culture, Sports, Science and Technology (MEXT) of Japan, MEXT KAKENHI Grant Nos. 22540327 and 26287062, by MEXT Elements Strategy Initiative to Form Core Research Center (TIES), and by the National Science Foundation under Grant No. NSF PHY11-25915 through Kavli Institute for Theoretical Physics, University of California at Santa Barbara.

\end{document}